\documentclass[12pt]{article}
\usepackage{epsfig,amssymb,longtable}

\newcommand{\eref}[1]{(\ref{#1})}

\newcommand{\cref}[1]{Chapter~\ref{#1}}
\newcommand{\bcenter}{\begin{center}}
\newcommand{\ecenter}{\end{center}}
\newcommand{\beq}{\begin{equation}}
\newcommand{\eeq}{\end{equation}}
\newcommand{\bea}{\begin{eqnarray}}
\newcommand{\eea}{\end{eqnarray}}
\newcommand{\bean}{\begin{eqnarray*}}
\newcommand{\eean}{\end{eqnarray*}}
\newcommand{\ba}{\begin{array}}
\newcommand{\ea}{\end{array}}
\newcommand{\ben}{\begin{enumerate}}
\newcommand{\een}{\end{enumerate}}
\newcommand{\bi}{\begin{itemize}}
\newcommand{\ei}{\end{itemize}}
\newcommand{\bd}{\begin{description}}
\newcommand{\ed}{\end{description}}
\newcommand{\bdiag}{\begin{diagram}}
\newcommand{\ediag}{\end{diagram}}

\def\IC{\mathbb{C}}

\def\IP{\mathbb{P}}
\def\IZ{\mathbb{Z}}
\def\cO{{\mathcal O}}

\newcommand{\BC}{\mathbb{C}}

\newcommand{\Fflat}{\mathcal{F}}
\newcommand{\tr}[1]{{\rm tr}[#1]}
\newcommand\cg{\mathfrak{g}}
\newcommand\CB{{\cal B}}
\newcommand\CF{{\cal F}}
\newcommand\CM{{\cal M}}
\newcommand\CN{{\cal N}}
\newcommand\CO{{\cal O}}
\newcommand\btheta{\overline{\theta}}
\newcommand\blambda{\overline{\lambda}}
\newcommand{\be}{\begin{equation}}
\newcommand{\ee}{\end{equation}}

\def\eps{\epsilon^{\alpha \beta}}
\def\barH{\overline{H}}

\newcommand{\drawsquare}[2]{\hbox{%
\rule{#2pt}{#1pt}\hskip-#2pt%
\rule{#1pt}{#2pt}\hskip-#1pt%
\rule[#1pt]{#1pt}{#2pt}}\rule[#1pt]{#2pt}{#2pt}\hskip-#2pt%
\rule{#2pt}{#1pt}}%
\newcommand{\fund}{\raisebox{-.5pt}{\drawsquare{6.5}{0.4}}}%
\newcommand{\antifund}{\overline{\fund}}

\begin{document}

\rightline{\small hep-th/0604208}
\rightline{\small DCPT-06/07}
\rightline{\small UPR-1151-T}

\vskip 0.5in
\centerline{\Large \bf Exploring the Vacuum Geometry of $\CN=1$ Gauge Theories}
\centerline{${}$}

\renewcommand{\thefootnote}{\fnsymbol{footnote}}

\centerline{{\bf
James Gray${}^{1}$\footnote{\tt gray@iap.fr},
Yang-Hui He${}^{2,3}$\footnote{\tt yang-hui.he@merton.ox.ac.uk},
Vishnu Jejjala${}^{4}$\footnote{\tt vishnu.jejjala@durham.ac.uk},
Brent D.\ Nelson${}^{5}$\footnote{\tt bnelson@sage.hep.upenn.edu}
}}
\begin{center}
${}^1${\it Institut d'Astrophysique de Paris and APC, Universit\'e de Paris 7, \\ 98 bis, Bd.\ Arago 75014, Paris, France}
\vskip 0.1in
${}^2${\it Merton College, Oxford University, \\ Oxford OX1 4JD, U.K.}
\vskip 0.1in
${}^3${\it Mathematical Institute, Oxford University, \\ 24-29 St.\ Giles', Oxford OX1 3LB, U.K.}
\vskip 0.1in
${}^4${\it Department of Mathematical Sciences, Durham University, \\ South Road, Durham DH1 3LE, U.K.}
\vskip 0.1in
${}^5${\it Department of Physics \& Astronomy, University of Pennsylvania, \\ 209 South 33rd St., Philadelphia, PA 19104, U.S.A.}
\end{center}

\setcounter{footnote}{0}
\renewcommand{\thefootnote}{\arabic{footnote}}

\begin{abstract}
Using techniques of algorithmic algebraic geometry, we present a new and efficient method for explicitly computing the vacuum space of $\CN=1$ gauge theories.
We emphasize the importance of finding special geometric properties of these spaces in connecting phenomenology to guiding principles descending from high-energy physics.
We exemplify the method by addressing various subsectors of the MSSM.
In particular the geometry of the vacuum space of electroweak theory is described in detail, with and without right-handed neutrinos.
We discuss the impact of our method on the search for evidence of underlying physics at a higher energy.
Finally we describe how our results can be used to rule out certain top-down constructions of electroweak physics.
\end{abstract}

\newpage

\section{Introduction and Summary}

It is %
often heard that definitive statements about high-energy physics cannot be obtained from low-energy experiments.
In particular, it is frequently maintained that one needs to achieve energy scales comparable to the natural scale of the theory under consideration in order to probe its validity.
This is not necessarily so.
Consider, for example, the case of supersymmetric grand unified field theories (GUTs).
These are well-defined, high-energy modifications of the Standard Model that have an associated energy scale of roughly $10^{16}$ GeV, below which the grand unified group breaks.
Yet, despite the large fundamental scale, the minimal $SU(5)$ version has been experimentally excluded by current limits on the lifetime of the proton~\cite{Murayama:2001ur}.
This is possible for the following reason:
once one adopts the organizing principle of high-energy grand unification certain testable properties of the low-energy theory immediately ensue.
The question we would like to pose is whether the paradigm can be inverted:
can the presence or absence of certain operators in the low-energy theory be used to ascertain guiding principles for the theory at high energies?
In other words, in the absence of experiments that directly probe the high-energy regime, it might simply suffice to develop a more clever low-energy attack tailored to the nature of the expected low-energy theory in order to infer physics at higher energies.

Of course this idea is among the oldest in particle physics.
The useful concept of 't Hooft naturalness~\cite{tHooft}, born as it is from our experience with gauge and discrete symmetry principles, has been a primary tool for low-energy model building for some time.
But in supersymmetric theories there are additional tools that we can utilize.
In this paper we expand upon our proposal~\cite{ghjn} to search for evidence of the nature of the high-energy completion of the Standard Model of particle physics within the context of the vacuum geometry of the low-energy theory itself.
While our search assumes nothing about the specifics of the true high-energy completion of phenomenological theories, it is especially sensitive to signals of underlying string physics.

The proposal is conceptually simple.
We wish to study the geometry of the vacuum space of the Minimal Supersymmetric Standard Model (MSSM).
The fundamental idea is that if we were to find that the geometry has some very special form, which
(a) is not and cannot be explained in terms of symmetries relating the relevant degrees of freedom in the effective field theory; and
(b) is astonishingly unlikely to have occurred by chance,
then this special form should be regarded as a consequence of as yet unknown physics.
For our purposes here, we shall take {\em special} to be synonymous with non-trivial properties of the algebraic geometry (such as the vanishing of certain topological numbers) describing the vacuum space.
The presence of any special geometrical structures in the vacuum space would be a collective consequence of such factors as the gauge group, the particle spectrum, and the interactions in the theory.
As such the special structures would tell us about the physics that determines these properties of our world.

Viewed from a bottom-up perspective, this proposal is more than simply a search for explanations of existing phenomenology.
It can be used to make predictions.
This is because the prior arguments lead very naturally to a new principle for constructing and extending low-energy phenomenological theories.
If a special geometry is found which is incredibly unlikely to have arisen by chance, then the existence of this geometry in the vacuum space should be taken as a fundamental property of the high-energy completion, whatever this happens to be.
As such when adding higher dimensional operators to our theory, only those compatible with this structure should be included.
This is a very restrictive constraint --- generically more restrictive than gauge invariance alone --- and therefore a highly predictive statement.
It means that any process mediated by those other higher dimensional operators should be suppressed.
Whether this is indeed true is something that may be tested experimentally.

Another application of a study of the vacuum space geometry of phenomenological theories is to rule out certain methods of top-down model building from ever achieving desired low-energy outcomes.
In some cases one of the most evident structures of a model created by a top-down construction is the geometry of its vacuum space.
A prime example of this is the growing industry of creating phenomenological theories on the worldvolume of a stack of D3-branes localized on a singularity.\footnote{
This program was initiated with~\cite{DM,KS,LNV}, with systematic progress in~\cite{Aldazabal:2000sa,BJL,wijnholt}.
Some algorithmic perspectives were developed in~\cite{HH,FHH}.}
In providing the same information for phenomenologically motivated theories themselves, the work presented here facilitates a comparison of the two.
Such a comparison can instantly rule out large classes of approaches to Standard Model building.
For example, if one wishes to construct a phenomenological theory which has a vacuum space which is {\em not} Calabi--Yau, then one cannot use a top-down approach which is known to always give rise to such a structure.

While the physical motivations for our studies are straightforward and easy to understand, new technical methods are required to study the geometry of the supersymmetric vacuum spaces that we encounter.
We make use of advances in algorithmic algebraic geometry to achieve our physical aims.
In due course, {\it we develop a new and efficient algorithm for computing the explicit vacuum moduli space of $\CN=1$ gauge theories.}
This involves the development of a novel method for determining the geometry of supersymmetric vacuum spaces, the solutions to the F- and D-flatness constraints of the theory.
For every solution to the F-terms, there exists a solution to the D-terms in the completion of the orbit of the complexified gauge group~\cite{lt}.
The D-orbits are themselves specified by the minimal set of holomorphic gauge invariant operators (GIOs) in the theory.
The algorithms we shall discuss in this paper apply the F-flatness constraints to the gauge invariant operators to obtain a description of the vacuum space as the solution to a set of algebraic relations among the operators.
One of the advantages of this attack is that it is built to exploit the computational algebro-geometric tools currently available.
As well, the methods we advance allow the careful study of the geometry of general $\CN=1$ theories in other contexts.
For this reason alone our investigation should be of interest to phenomenologists, formal theorists, and mathematicians alike.

Though our methodology certainly applies to the full MSSM, it is unfortunately the case that the number and complexity of the gauge invariant operators renders this field theory too complicated to analyze with standard computational means.
Therefore, while the use of a supercomputer to attack the full problem is clearly a project of great importance --- and one that we are currently pursuing --- we will in this paper concentrate on subsectors of particular interest for the sake of illustrating and applying the methodology.
Already, there is interesting physics here.
In particular, we find that the existence of non-trivial geometry in the vacuum space of the MSSM depends crucially on the number of generations of matter fields.
Restricting to the electroweak sector by setting vacuum expectation values (vevs) of quarks to zero by hand,
we find that special structures emerge.
When dimension four R-parity preserving terms are added to the renormalizable superpotential with $\mu$-term and standard Yukawa interactions, the moduli space of vacua is an affine cone over the Veronese surface.
The Veronese surface is one of the simplest varieties encountered in algebraic geometry.
Adding R-parity violating operators destroys the structure entirely:
the geometry of the vacuum space becomes trivial, being either a line or a point.
Similarly, when we include a right-handed neutrino with Majorana and Dirac mass terms, the vacuum space is again a cone over a Veronese base, only this time at the renormalizable level itself.
This structure is stable under R-parity preserving interactions at higher mass level.
These results suggest that geometry and phenomenology might well be correlated.
We have carried out a general survey of superpotentials in the electroweak sector to discover how special these structures are.

The outline of this paper is as follows.
In Section 2, we detail our methodology for computing the moduli space as an algebraic variety.
We include a number of examples to illustrate our techniques explicitly.
In Section 3, we discuss the moduli space of the MSSM.
As we have indicated, we will focus much of our attention on supersymmetric electroweak theory.
This is both a computationally tractable problem and an interesting laboratory for finding and mapping special geometries.
In Section 4, we discuss prospects for continuing this program.
A short Appendix explains the relevant algebraic geometry to physicists who may be initially unfamiliar with the language.

\section{Computing Moduli Spaces of ${\cal N}=1$ Gauge Theories}

In this Section, we begin by reviewing the method of Luty and Taylor~\cite{lt} for describing the moduli space of $\CN=1$ supersymmetric gauge theories.
We then proceed to describe how this perspective on the geometry of the space of supersymmetric vacua is natural for algebraic geometry.
The vacuum space of the gauge theory is captured as the image of a particular ring map.
In this form the geometry can be efficiently investigated algorithmically.
We illustrate our discussion with simple examples.

\subsection{The Moduli Space}

Consider the $\CN=1$ globally supersymmetric theory defined by
\be
\label{action}
S = \int d^4x\ \left[ \int d^4\theta\ \Phi_i^\dagger e^V \Phi_i +
    \left( \frac{1}{4g^2} \int d^2\theta\ \tr{W_\alpha W^\alpha} +
    \int d^2\theta\ W(\Phi) + {\rm h.c.} \right) \right].
\ee
The $\Phi_i$ are chiral superfields transforming in some representation $R_i$ of the (compact) gauge group $G$; $V$ is a vector superfield transforming in the Lie algebra $\cg$; $W_\alpha = i\overline{D}^2 e^{-V} D_\alpha e^V$, the gauge field strength, is a chiral spinor superfield; and $W(\Phi)$ is the superpotential, which is a holomorphic function of the $\Phi_i$.

The standard discussion, as found in~\cite{wessandbagger} for example, tells us that field configurations in global superspace
which do not break supersymmetry extremize the superpotential when treated as a function of the scalar components of the chiral fields:
\be
\left. \frac{\partial W(\phi)}{\partial \phi_i}
\right|_{\phi_i=\phi_{i 0}} = 0. \label{fterm} \ee
Here the $\phi_{i 0}$ are vacuum expectation values of the scalar components of $\Phi_i$.
These are the {\em F-flatness} conditions.
For each generator of $G$, there is also a {\em D-flatness} condition that supersymmetric field configurations must satisfy.
In Wess--Zumino gauge, this condition takes the familiar form
\be D^A = \sum_i \phi_{i 0}^\dagger\, T^A\, \phi_{i 0} = 0,
\label{dterm} \ee
where the $T^A$ are the generators of the gauge group.

In principle, we could examine the vacuum space of gauge theories just by restricting to the subspace of field configurations in the theory that satisfy both the F- and D-flatness constraints.
Removing the gauge redundancy within this subspace then returns to us the space of interest.
However, for complicated examples such as the MSSM this procedure makes studying the nature of the resulting geometry calculationally intensive.
We shall instead obtain the supersymmetric vacuum space in a different manner, one that is designed to give us the geometry in a form that is then amenable to an analysis of its properties on a computer.

The first step in reformulating the description of the moduli space is to switch from Wess--Zumino gauge to a less restrictive gauge.
The action~\eref{action} has a huge gauge invariance that we do not normally see.
This invariance may be parameterized by a chiral superfield gauge parameter $\Lambda\in\cg$:
\bea \label{DWZ}
\Phi \mapsto g\cdot \Phi, && e^V \mapsto g^{\dagger\ -1} e^V g^{-1},
\eea
where $g = e^{i\Lambda}$.
In particular, this chiral superfield parameter can be taken to be a complex scalar, giving a subset of the invariance of the action which is the complexification $G^c$ of the original gauge group $G$.

Luty and Taylor~\cite{lt} choose a gauge where the residual gauge invariance left over from the full symmetry of the action is precisely this complexification.
In this gauge the vector superfield has an expansion
\be
V_A = C_A - \theta\sigma^\mu\btheta v_{\mu A} + i\theta\theta\btheta\blambda_A - i\btheta\btheta\theta\lambda_A + i\theta\theta\btheta\btheta D_A, \label{va}
\ee
which can alternatively be taken as the definition of the gauge.
The right hand side of~\eref{va} differs from Wess--Zumino gauge because of the addition of the degree of freedom $C_A$.

We wish to study the space of supersymmetric configurations of the theory.
The F-term constraints are unchanged from their usual expressions in this gauge.
Imagine that we take a solution to the F-flatness conditions, which we shall again denote by $\phi_{i 0}$.
The D-term constraints in this gauge are given by the equations
\be \label{DNWG} \frac{\partial}{\partial C_A} \sum_i \phi_{i 0}^\dagger e^C \phi_{i 0}=0,
\ee
where $C = C_A T^A$.
Substituting the solution $\phi_{i 0}$ from the F-flatness conditions into the above D-term conditions yield equations for $C_{A 0}$.
The F-terms are holomorphic quantities in the fields that carry a gauge index and thus are covariant under changes of the imaginary part of the gauge group.
Consequently, we can perform such a transformation on our solution to the F-term constraints and recover another solution.
We use this freedom to rotate the solution for $C_{A 0}$ to the D-terms to the point where $C = 0$.
When we do this, we of course obtain a solution to the D-terms in the standard Wess--Zumino gauge, which we have regained.
(That the correct D-terms are obtained in this gauge can be checked by comparing~\eref{dterm} to~\eref{DNWG} with $C_A$ set to zero once we have differentiated.)
The non-trivial statement here is that for every solution to the F-term constraints there is \emph{one and only one} solution to the D-flatness conditions in Wess--Zumino gauge.
In this sense the D-flatness conditions are simply a gauge fixing condition.
For a more careful analysis of this and related points we refer the reader to the original literature~\cite{lt,Buccella:1982nx,Gatto:1986bt,Procesi:hr,witten93}.

In the end, the relevant observation for us is that the space of all supersymmetric vacua is given by the space of all of the solutions to the F-term constraints $\CF$ modulo the complexification of the gauge group (the real part being used to remove the standard gauge redundancy and the imaginary part to obtain the solution to the D-terms in Wess--Zumino gauge).
The moduli space is the symplectic quotient:
\be
\CM = \CF//G^c.
\ee
Moreover, the moduli space of vacua is an {\em algebraic variety}.\footnote{
The definition of an {\em algebraic variety} is given in the Appendix to this paper.}

\subsection{Parameterizing the Moduli Space}

We have seen that for any solution to the F-terms, there also exists a solution to the D-terms in the completion of the orbit of the complexified gauge group.
It is intuitively reasonable that the moduli space $\CM$ will be parameterized by the set of holomorphic gauge invariant operators, with relations among them.
We generally expect such relations to exist because there will generically be more gauge invariant operators for any particular theory than dimensions in the vacuum space.
That is to say, the gauge invariant operators are an over-complete set for describing F- and D-flat field directions.
The D-orbits are the loci in the space of fields, described in the previous section, for which the D-terms are stationary with respect to $G^c$.
Because holomorphic gauge invariant operators are invariant under $G^c$, they provide a basis for labeling these orbits.
Additional relations among gauge invariant operators arise because they are built out of fields and these fields must themselves satisfy relations --- the F-term constraints.

Luty and Taylor prove the following theorem~\cite{lt}:
\begin{quotation}
{\bf Theorem} Given a group $G^c$ acting on a variety $A$, there is a one-to-one correspondence between $A//G^c$ and the set of points in the affine variety $A^G$ defined by the ring $R_G$ of $G$-invariant elements in $R=R(A)$, where $R(A)$ is the ring of polynomials defining the variety $A$.
\end{quotation}
The algebro-geometric concepts in the statement of the theorem and in the discussion that follows are disentangled for the reader in the Appendix of our paper.
What the statement means for our purposes is that the vacuum space $\CM = \CF//G^c$ can be described in algebraic geometry in terms of the ring consisting of all polynomials built out of holomorphic gauge invariant operators that are themselves built from constant field configurations, which are solutions to the F-term equations.

\subsubsection{The Computational Problem}
In a language conducive to calculation, the issue with which we are confronted is the following problem in polynomial computation.
Given an $\CN=1$ theory, with superfields whose scalar components are $\{ \phi_i \}$ and superpotential $W(\Phi_i)$ ($i=1, \ldots, n$), one always has a finite generating set, $D := \{r_j(\phi_i) \}$, of gauge invariant operators encoding the D-terms ($j = 1, \ldots, k$).
We here suppress gauge indices for the fields for convenience, but in the actual computation we will expand everything into components.
Any gauge singlet is then a polynomial in the elements of $D$, which is to say that the elements of $D$ generate the chiral ring of gauge invariant operators.
The gauge invariant operators in the set $D$ are themselves polynomials in the (components) of the fields $\phi_i$.
The vacuum moduli space will be parameterized by this set.
That is, the coordinates of the vacuum space $\CM$ as an algebraic variety are the $r_j$.

In the case when the superpotential is vanishing, the F-terms are trivial, and the equations of $\CM$ are simply the (non-trivial, independent) algebraic relations amongst the elements of $D$.
To determine these relations is a standard {\em syzygy problem}.\footnote{
The {\em syzygy problem} is simply to find relations among generators of an ideal.
The word {\em syzygy} is borrowed from astronomy where it denotes an alignment (conjunction or opposition) of celestial bodies.}
Let there be $d$ such independent relations;
that is, there are $d$ equations in the $k$ variables $r_j$.
Then the moduli space is a variety defined by the $d$ equations in $\IC[r_1, \ldots, r_k]$.
In the case of $n$ F-terms $\{f_i = \partial_i W = 0 \}$, one needs to find all relations subject to the F-term constraints.

It is expedient to rephrase the above discussion in terms of an algorithm for formulating the space in the language of computational algebraic geometry.
A systematic attack can then be directed employing such computer packages such as {\tt Macaulay~2}~\cite{mac} or {\tt Singular}~\cite{sing}.
We make extensive use of this technology in performing computations throughout this paper.

\subsubsection{The Algorithm} \label{s:algo}
\begin{enumerate}

\item
The $n$ fields $\{\phi_1,\ldots,\phi_n\}$ define the ring $R := \IC[\phi_1,\ldots,\phi_n]$.
Elements of this ring are polynomials in the fields $\phi_i$.
In general, a monomial expression $\phi_1^{m_1} \cdots \phi_n^{m_n}$ will transform non-trivially under the action of the gauge group $G$.
The F-term constraints are $\partial_i W = 0$.
Since there are $n$ fields, there will be $n$ (possibly trivial) F-flat conditions, each coming from taking the derivative of the superpotential with respect to a field and each of the form of a polynomial (in the $n$ variables $\{\phi_i\}$) being set to zero.
Note that the F-terms will also in general carry a representation of the gauge group.
The key is that the F-terms furnish an ideal $F := \langle \langle f_1, \ldots, f_n \rangle \rangle$ of the polynomial ring $R$, where each $f_i = \partial_i W$ is a polynomial F-flatness equation.
(Double angle brackets are used to indicate the generators of the ideal.)
By definition, then, the quotient ring $\IC[\phi_1, \ldots, \phi_n] / F$ is a polynomial ring in which all F-flat conditions $f_i = 0$ are satisfied.
Therefore, F-flatness is automatically imposed by working in the quotient ring
\beq\label{Fflat}
\Fflat = R / \langle\langle \partial_i W \rangle\rangle.
\eeq
In practice, one always puts the generators of the ideal of F-terms in standard Gr\"obner basis.

\item
The physics in the D-terms are exactly captured by holomorphic gauge invariant operators as these are constant along orbits in $G^c$.
We take a minimal generating set of gauge invariant operators $D = \{ r_j(\{\phi_i\}) \}$, where $j=1,\ldots,k$.
As the generators $r_j$ are polynomial in the fields $\phi_i$, $D$ can be regarded as defining a ring map from $R$ to $S := \IC[r_1, \ldots, r_k]$.
Again, the $r_j$ are minimally placed in a Gr\"obner basis.
In the absence of the F-terms, the syzygies of this map will give the $d$ independent relations amongst the $r_k$.
To impose F-term constraints, one simply regards $D$ as a map from the quotient ring $\Fflat$ above where F-flatness is automatic.
That is, we have a map from $\Fflat$ to $S$ as
\beq\label{giomap-noW}
\Fflat = \IC[\phi_1, \ldots, \phi_n] / \langle\langle \partial_i W \rangle\rangle \stackrel{D}{\longrightarrow} \IC[r_1, \ldots, r_k].
\eeq

\item
The moduli space of F- and D-flat configurations is the image ${\rm Im}(D)$ of this map.
It is the space of all holomorphic gauge invariant operators built out of F-flat field configurations.
The vacuum manifold
\be
\CM \simeq {\rm Im}\left(\Fflat \stackrel{D}{\longrightarrow} S
\right)
\ee
is an ideal of $S = \IC[r_1, \ldots, r_k]$ and is an affine variety in $\IC^k$.
\end{enumerate}

Once we have the geometry in the form of an image of a ring map in this manner, we use {\tt Macaulay~2} and {\tt Singular} to study its properties algorithmically.
These programs manipulate the types of polynomial systems described above using Gr\"obner basis techniques.
A good introductory guide to these kinds of methods can be found in~\cite{cox}.
A clear advantage of the algorithm outlined here is that the {\it explicit} affine equation of $\CM$ can be given as the intersection of polynomial equations.
\subsection{Some Examples}\label{s:eg}

We shall now provide some simple examples to illustrate how the prescription works in practice.
As well as serving as a demonstration of our methods, we shall see that these examples correctly reproduce known results.
In addition, we have selected these cases to show how special geometry can indeed arise in phenomenological compactifications of superstring theory.
This validates the particular power of our method in searching for evidence of high-energy completions or in ruling out certain models.

\subsubsection{The Conifold}
We shall first illustrate our procedure by the famous D3-brane worldvolume theory at a conifold singularity~\cite{conifold}.
Consider placing a stack of $N$ parallel D3-branes on the surface containing the conifold singularity defined as a hypersurface in $\IC^4$:
\beq\label{coni}
\{ u v - z w = 0 \} \subset \IC[u,v,z,w].
\eeq
The result is a quiver theory~\cite{DM} with two nodes as depicted below:
\beq
\ba{cc}
\ba{ccc}
& SU(N) & SU(N) \\
x_{i=1,2} & \fund & \antifund \\
y_{j=1,2} & \antifund & \fund
\ea
&
\ba{r}
\epsfxsize = 3cm\epsfbox{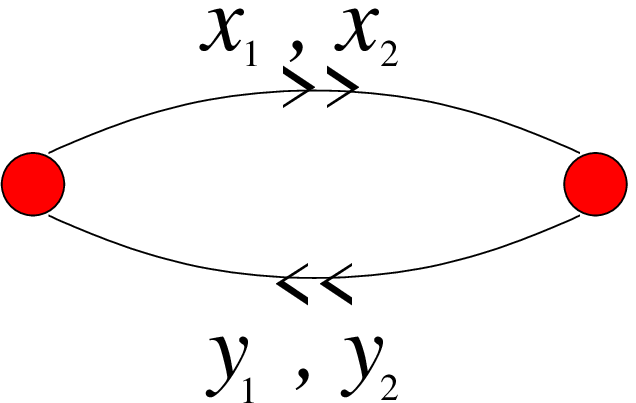}
\ea
\ea
\eeq
It is an $SU(N) \times SU(N)$ theory with fields $(x_1, x_2)$ charged as $(\fund, \antifund)$ (in standard Young Tableaux notation) and $(y_1, y_2)$ as $(\antifund, \fund)$ together with a superpotential
\beq
W = {\rm Tr}(x_1 y_1 x_2 y_2 - x_1 y_2 x_2 y_1).
\eeq
Now, for the simple case of a single brane, {\em i.e.}\ $N=1$, $W$ clearly vanishes as the matrices $x_i,y_i$ now commute.
Then, the supersymmetric vacuum space of the theory is generated by the four polynomials
\be
\{ r_1, r_2, r_3, r_4 \} = \{ x_i y_j \}_{i,j=1,2}
\ee
without further constraints.
The moduli space $\CM$ is hence given by the relations among the $r_i$.
In this case there is a single relation and if we define
\beq
u = x_1 y_1, \ \ v = x_2 y_2, \ \ z = x_1 y_2, \ \ w = x_2 y_1,
\eeq then the relation is precisely~\eref{coni}.
This must be so by construction.
Since the degrees of freedom in the $\CN=1$ worldvolume theory correspond to the position moduli of the brane, the moduli space of the D-brane probe is the local equation of the Calabi--Yau space containing the singularity.

In terms of our three-step algorithm, the above procedure is realized as follows.
Here, $R = \IC[x_1, x_2, y_1, y_2]$, $D = \{ r_1, r_2, r_3, r_4 \} = \{u = x_1 y_1, v = x_2 y_2, z = x_1 y_2, w  = x_2 y_1\}$ and $S = \IC[u,v,z,w]$.
The first step in this case is trivial because we have no F-term constraints.
Thus the ring of polynomials describing the space of F-flat conditions is simply the same as that describing field space itself;
it is thus the ring $R$ of all polynomials in the fields.
The map $D$ takes $R$ to $S$ with a single relation $uv = zw$.

\subsubsection{Non-trivial F-terms}\label{s:z3}

The case above for $N>1$ is slightly more involved as we must add non-trivial F-terms constraints;
the explicit parameterization can be found, for example, in~\cite{Berenstein:2001uv}.
Let us illustrate the case of non-trivial F-terms instead with another well-known example, a D3-brane on the orbifold $\IC^3/\IZ_3$ with action $(1,1,1)$.
The gauge theory is a $U(1)^3$ quiver theory:
\beq
\ba{cc}
\ba{l}
\epsfxsize = 3cm\epsfbox{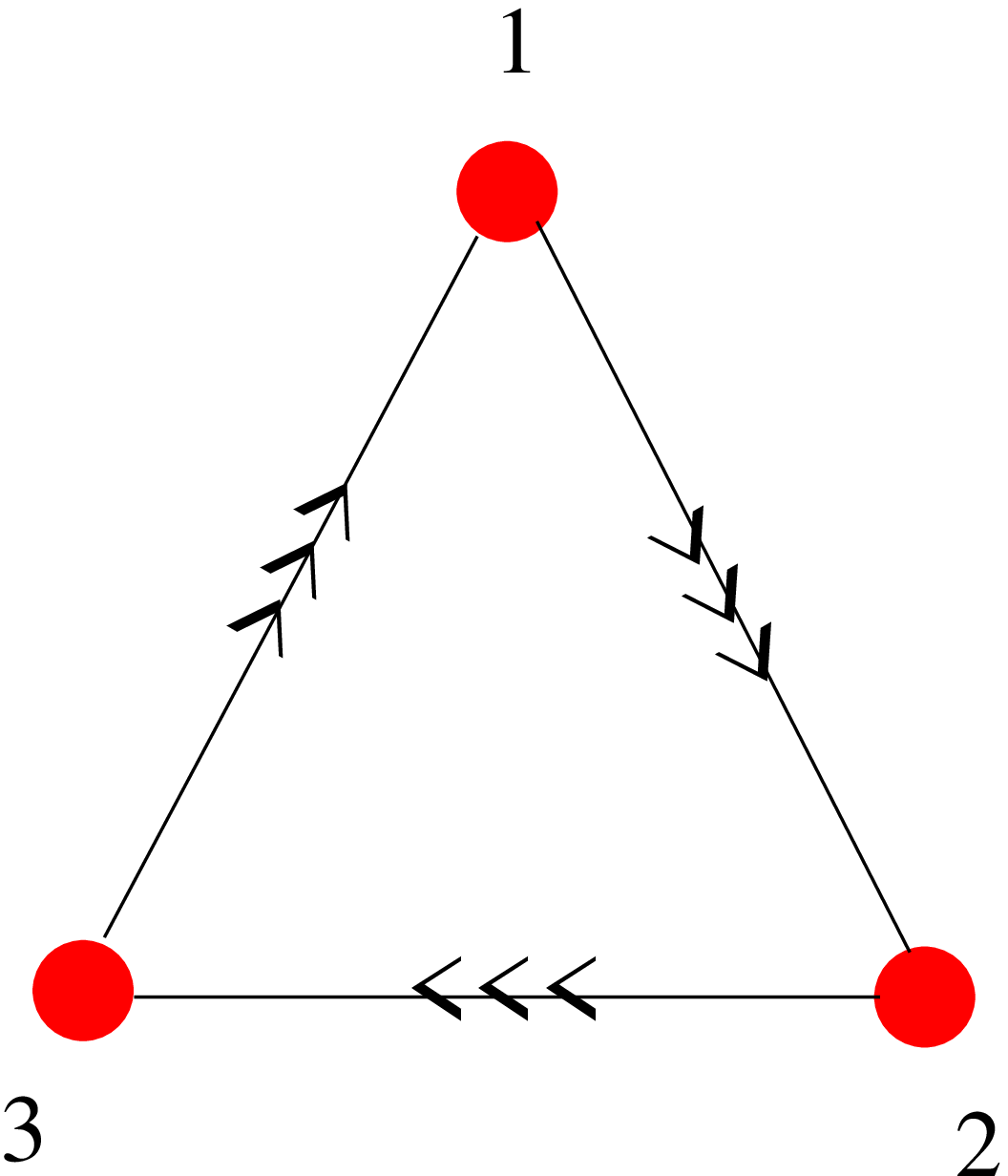}
\ea
&
W=\epsilon_{\alpha\beta\gamma} X^{(\alpha)}_{12} X^{(\beta)}_{23} X^{(\gamma)}_{31}.
\ea
\eeq
There are nine fields $X^{(\alpha)}_{12}, X^{(\beta)}_{23}$, and $X^{(\gamma)}_{31}$, $\alpha,\beta,\gamma=1,2,3$ and $27$ gauge invariant operators formed by their products corresponding to the closed cycles in the quiver.
Note that this is an example in which the superpotential consists of all possible gauge-invariant operators at the renormalizable level.
The moduli space is
\beq
\CM \simeq {\rm Im} \left( \frac{\IC[X^{(\alpha)}_{12}, X^{(\beta)}_{23},X^{(\gamma)}_{31}]}
                                {\langle\langle \epsilon_{\alpha\beta\gamma} X^{(\beta)}_{23}X^{(\gamma)}_{31},
                                                \epsilon_{\alpha\beta\gamma} X^{(\alpha)}_{12}X^{(\gamma)}_{31},
                                                \epsilon_{\alpha\beta\gamma} X^{(\alpha)}_{12}X^{(\beta)}_{23} \rangle\rangle}
                           \stackrel{D = \{X^{(\alpha)}_{12} X^{(\beta)}_{23}X^{(\gamma)}_{31} \}}{\longrightarrow}
                           \IC[r_1, \ldots, r_{27}] \right).
\eeq
We find, using~\cite{mac}, that the image of the map $D$ is an affine variety of dimension three and degree nine;
it is the intersection of $17$ lines and $27$ quadratics in $\IC^{27}$.
The presentation of these $44$ equations is uninstructive.
However, we shall set the notation that $(k | d, \delta | m_1^{n_1} m_2^{n_2} ... )$ signifies a variety of complex dimension $d$ and degree $\delta$, given as the intersection of $n_i$ polynomials, each of degree $m_i$ in $\IC^k$.
For the cases we shall present in this paper, the affine varieties are intersections of homogeneous polynomials.
As such, we can write them as affine cones over a compact projective variety of one lower dimension. 
Hence,
\bea \label{eq:ac}
(k | d, \delta | m_1^{n_1} m_2^{n_2} ... ) &:=&
\mbox{Affine variety of complex dimension $d$, realized as an affine} \\
&&\mbox{cone over a projective variety of dimension $d-1$  and degree $\delta$,} \nonumber \\
&&\mbox{given as the intersection of $n_i$ polynomials of degree $m_i$ in $\IP^k$.} \nonumber
\eea

In this notation, the moduli space for the above theory is $(27 | 3, 9 | 1^{17} 2^{27})$.
Now, the $17$ linear relations are trivial and allow us to eliminate $17$ variables, leaving us with only $10$.
The result is actually that $\CM = (10 | 3, 9 | 2^{27})$ and is in fact given by the $27$ quadratic relations amongst the (affine cone over the) standard cubic (Veronese) embedding:
\be
\ba{ccc}
\IP^{2} & \rightarrow & \IP^{9} \cr
[x_0:x_1:x_2] & \mapsto &
[{x_0}^3 : {x_0}^2x_1 : x_0{x_1}^2: {x_1}^3: {x_0}^2x_2 : x_0x_1x_2: {x_1}^2 x_2:  x_0{x_2}^2: x_1{x_2}^2:{x_2}^3].\label{P2P9}
\ea
\ee
Affinizing this to a map from $\IC[x_0, x_1, x_2]$ to $\IC[y_1, \ldots, y_{10}]$, where the $y_i$ are the $10$ cubic terms on the right hand side of~(\ref{P2P9}), we see that the image is an affine variety in $\IC^{10}$, with the relations
\beq
\ba{llllll}
\{
{y_2}^2 - y_1y_3, & y_2y_3 - y_1y_4, & {y_3}^2 - y_2y_4, & y_2y_5 - y_1y_6, & y_3y_5 - y_1y_7, & y_2y_6 - y_1y_7, \\
y_4y_5 - y_2y_7, & y_3y_6 - y_2y_7, & y_4y_6 - y_3y_7, & {y_5}^2 - y_1y_8, & y_5y_6 - y_1y_9, & y_2y_8 - y_1y_9, \\
{y_6}^2 - y_2y_9, & y_5y_7 - y_2y_9, & y_3y_8 - y_2y_9, & y_6y_7 - y_3y_9, & y_4y_8 - y_3y_9, & {y_7}^2 - y_4y_9, \\
y_5y_8 - y_1y_{10}, & y_6y_8 - y_2y_{10}, & y_5y_9 - y_2y_{10}, & y_7y_8 - y_3y_{10}, & y_6y_9 - y_3y_{10}, & y_7y_9 - y_4y_{10}, \\
{y_8}^2 - y_5y_{10}, & y_8y_9 - y_6y_{10}, & {y_9}^2 - y_7y_{10}\}. & & &
\ea
\eeq
This defines the three-dimensional orbifold $\IC^3/\IZ_3$, as specified by the total space of $\CO_{\IP^2}(-3)$.
(The reader is referred to the Appendix for this standard notation for line bundles).
Of course, this is no surprise to us~\cite{DM,dgm}.
The moduli space, in a D-brane probe scenario, must locally be the singularity probed {\it ab initio}.
This is then a good check of our method.

\section{The MSSM Moduli Space}

The preceding section has provided a general methodology for efficiently computing properties of the vacuum space of $\mathcal{N}=1$ supersymmetric theories.
Clearly the most important supersymmetric theory is the minimal supersymmetric extension of the Standard Model (MSSM).
The existence of supersymmetry at the TeV scale will be determined at the LHC, as will the existence of supersymmetric representations of the Standard Model gauge group.
For the remainder of this article we will devote our attention exclusively to this class of models.

\subsection{The Full Theory}

The MSSM is an $\CN=1$ globally supersymmetric gauge theory in four dimensions.
As such our method can be applied here in exactly the same way as it was for the examples in the previous section.
An earlier systematic classification of the flat directions in the MSSM was undertaken in the pioneering work of~\cite{mssm};
we shall largely adhere to their nomenclature.
We emphasize, however, that our method goes well beyond finding flat directions and the dimensions of the moduli space ---
it finds the vacuum space explicitly as an algebraic variety.

The field content of the MSSM constitute the input data.
This is given in Table~\ref{tbl:MSSM} below, along with our index conventions.
Once we expand into components, there are a total of $18+9+9+6+3+2+2 = 49$ scalars.
This means we are initially working over the polynomial ring $\IC[\phi_{1}, \ldots, \phi_{49}]$, which is significantly larger than the toy examples presented in Section~\ref{s:eg}.

\begin{table}[thb]
{\begin{center}
\begin{tabular}{ccc}
\begin{tabular}{|c|c|}\hline
INDICES & \\ \hline \hline
$i,j,k,l = 1,2,3$ & Flavor (family) indices \\
$a,b,c, d = 1,2,3$ & $SU(3)_C$ color indices \\
$\alpha, \beta, \gamma, \delta = 1,2 $ & $SU(2)_L$ indices \\
\hline
\end{tabular}
&
\begin{tabular}{|c|c|}\hline
FIELDS & \\ \hline \hline
$Q_{a, \alpha}^i$ & $SU(2)_L$ doublet quarks \\
$u_a^i$ & $SU(2)_L$ singlet up-quarks \\
$d_a^i$ & $SU(2)_L$ singlet down-quarks \\
$L_{\alpha}^i$  & $SU(2)_L$ doublet leptons \\
$e^i$ & $SU(2)_L$ singlet leptons \\
$H_\alpha$ & up-type Higgs \\
$\barH_\alpha$ & down-type Higgs \\ \hline
\end{tabular}
\end{tabular}
\end{center}}
{\caption{\label{tbl:MSSM} {\bf Index conventions and field content
of the MSSM}. }}
\end{table}
The first step in applying our method to the MSSM is to look at the submanifold of the field space spanned by the above fields which corresponds to F-flat configurations.
To work out the F-flatness constraints we require the superpotential.
We could for example take the following ``minimal'' renormalizable expression
\bea\label{renorm}
W_{\rm minimal} &=& C^0 \sum_{\alpha, \beta} H_\alpha \barH_\beta \eps + \sum_{i,j} C^1_{ij} \sum_{\alpha, \beta, a} Q^i_{a,\alpha} u^j_a H_\beta \eps \nonumber \\
&& +\sum_{i,j} C^2_{ij} \sum_{\alpha, \beta, a} Q^i_{a,\alpha} d^j_a \barH_\beta \eps + \sum_{i,j} C^3_{ij} e^i \sum_{\alpha, \beta} L^j_{\alpha} \barH_\beta \eps.
\eea
Here $C^0$, $C^{1,2,3}_{ij}$ are flavor mixing matrices and are constant coefficients in the superpotential;
there is a total of $1 + 3 \cdot 3^2 = 28$ of these parameters.
In general these are complex parameters, but for the purpose of obtaining topological information about the vacuum manifold defined by~(\ref{renorm}) we can take these numbers to be real without loss of generality.
In fact, we can restrict the entries of these matrices to be integer valued for ease of computation.\footnote{
We choose these integer values at random prior to performing the computations.
To ensure that this restriction does not lead to accidental cancelations among various terms within the F-flatness conditions, we then repeat the calculation with a different random number seed.}
Indeed, the particular values of the coefficients determine the complex structure of the vacuum variety.
Finally, dimension counting reveals the matrices $C^{1,2,3}_{ij}$ to be dimensionless while $C^0$ carries positive mass dimension.
This last coefficient is, of course, the $\mu$-parameter of the MSSM superpotential.
As we are uninterested in the scale of vevs along flat directions we can absorb the mass parameter into the coefficient as we have done in~(\ref{renorm}).

The F-term constraints can be obtained from the above superpotential by setting its derivatives with respect to the fields to zero.
Once we have the F-flatness conditions we can create the quotient ring describing the geometry of F-flat configuration space exactly as we did in the examples in the previous section.
Here this is determined by a list of $49$ polynomials in~$R$ which generate the ideal $F = \langle \langle \partial_i W \rangle \rangle$.
We then must find the quotient ring $\Fflat = R / F$ as defined in~\eref{Fflat}.

The second of step for applying our algorithm to the MSSM involves an investigation of the holomorphic gauge invariant operators in the theory.
Gherghetta, Kolda, and Martin have performed this analysis;
according to their paper~\cite{mssm}, the space of gauge invariant operators of the MSSM is generated by a finite set of $28$ types.
These are listed in Table~\ref{tbl:gio}, where we define
\beq
(QQQ)_4 := [(QQQ)_4]_{\alpha\beta\gamma} = Q^i_{a, \alpha} Q^j_{b, \beta} Q^k_{c, \gamma} \epsilon^{abc}\epsilon^{ijk},
\eeq
which transforms as a singlet of $SU(3)_C$ and as a $\mathbf{4}$ of $SU(2)_L$,
and where the notation $\mbox{antisymmetric}\{(i,j),(k,m)\}$ means that the multi-indices $I = (i,j)$ and $J = (k,m)$ for $i,j,k,m=1,2,3$ are antisymmetrized so that $I,J = 1, \ldots, 9$ and $J < I$.
As can be seen from this table, the field content and gauge group structure of the MSSM is considerably more complicated than any of the previous examples from Section~\ref{s:eg}.
This results in many more gauge invariant operators, each of which having an appreciably more involved structure.
We count a total of $991$ operators.
That is, the map $D = \{r_i\}$ maps $\Fflat \subset \IC[\phi_1,\ldots,\phi_{49}]$ to $S = \IC[r_1, \ldots, r_{991}]$.
The image of the ring map from the ring describing the F-flat configurations to the ring $S$ is then the vacuum space we desire.

\begin{table}
{\begin{center}\begin{footnotesize}
\begin{tabular}{|c||c|c|c|}\hline
\mbox{Type} & \mbox{Explicit Sum} & \mbox{Index} & \mbox{Number} \\
\hline \hline
$LH$  & $L^i_\alpha H_\beta \eps$ & $i=1,2,3$ & 3 \\ \hline
$H\barH$ & $H_\alpha \barH_\beta \eps$ & & 1  \\ \hline
$udd$ & $u^i_a d^j_b d^k_c \epsilon^{abc}$ & $i,j=1,2,3$; $k=1,\ldots,j-1$ & 9  \\ \hline
$LLe$ & $L^i_\alpha L^j_\beta e^k \eps$ & $i,j=1,2,3$; $k=1,\ldots,j-1$ & 9  \\ \hline
$QdL$ & $Q^i_{a, \alpha} d^j_a L^k_\beta \eps$ & $i,j,k=1,2,3$ & 27 \\ \hline
$QuH$ & $Q^i_{a, \alpha} u^j_a H_\beta \eps$ & $i,j=1,2,3$ & 9 \\ \hline
$Qd\barH$ & $Q^i_{a, \alpha} d^j_a \barH_\beta \eps$ & $i,j=1,2,3$ & 9 \\ \hline
$L\barH e$ & $L^i_\alpha \barH_\beta \eps e^j$ & $i,j=1,2,3$ & 9 \\ \hline
$QQQL$ & $Q^i_{a, \beta} Q^j_{b, \gamma} Q^k_{c, \alpha} L^l_\delta \epsilon^{abc} \epsilon^{\beta\gamma}\epsilon^{\alpha\delta}$ & $\ba{l} i,j,k,l=1,2,3; i\ne k, j\ne k, \\ j<i, (i,j,k) \ne (3,2,1) \ea$ & 24 \\ \hline
$QuQd$ & $Q^i_{a, \alpha} u^j_a Q^k_{b, \beta} d^l_b \eps$ & $i,j,k,l=1,2,3$ & 81 \\ \hline
$QuLe$ & $Q^i_{a, \alpha} u^j_a L^k_{\beta} e^l \eps$ & $i,j,k,l=1,2,3$ & 81 \\ \hline
$uude$ & $u^i_a u^j_b d^k_c e^l \epsilon^{abc}$ & $i,j,k,l=1,2,3; j<i$ & 27 \\ \hline
$QQQ\barH$ & $Q^i_{a, \beta} Q^j_{b, \gamma} Q^k_{c, \alpha} \barH_\delta \epsilon^{abc} \epsilon^{\beta\gamma} \epsilon^{\alpha\delta}$ & $\ba{l} i,j,k,l=1,2,3; i\ne k, j\ne k, \\ j<i, (i,j,k) \ne (3,2,1) \ea$ & 8 \\ \hline
$Qu\barH e$ & $Q^i_{a, \alpha} u^j_a \barH_\beta e^k \eps$ & $i,j,k =1,2,3$ & 27 \\ \hline
$dddLL$ & $d^i_a d^j_b d^k_c L^m_\alpha L^n_\beta \epsilon^{abc} \epsilon_{ijk} \eps$ & $m,n=1,2,3, n<m$ & 3 \\ \hline
$uuuee$ & $u^i_a u^j_b u^k_c e^m e^n \epsilon^{abc} \epsilon_{ijk}$ & $m,n=1,2,3, n \le m$ & 6 \\ \hline
$QuQue$ & $Q^i_{a, \alpha} u^j_a Q^k_{b, \beta} u^m_b e^n \eps$ & $ \ba{l} i,j,k,m,n=1,2,3; \\ \mbox{antisymmetric}\{ (i,j), (k,m) \}
\ea$ & 108  \\ \hline
$QQQQu$ & $Q^i_{a, \beta} Q^j_{b, \gamma} Q^k_{c, \alpha} Q^m_{f,\delta} u^n_f \epsilon^{abc} \epsilon^{\beta\gamma} \epsilon^{\alpha\delta}$ & $\ba{l} i,j,k,m=1,2,3; i\ne m, j\ne m, \\ j<i, (i,j,k) \ne (3,2,1) \ea$ & 72  \\ \hline
$dddL\barH$ & $d^i_a d^j_b d^k_c L^m_\alpha \barH_{\beta} \epsilon^{abc}\epsilon_{ijk} \eps$ & $m=1,2,3$ & 3 \\ \hline
$uudQdH$ & $u^i_a u^j_b d^k_c Q^m_{f, \alpha}d^n_f H_\beta \epsilon^{abc} \eps$ & $i,j,k,m=1,2,3; j<i$ & 81 \\ \hline
$(QQQ)_4LLH$  & $(QQQ)_4^{\alpha\beta\gamma} L^m_\alpha L^n_\beta H_\gamma$ & $m,n=1,2,3, n<=m$ & 6 \\ \hline
$(QQQ)_4LH\barH$ & $(QQQ)_4^{\alpha\beta\gamma} L^m_\alpha H_\beta \barH_\gamma$ & $m=1,2,3$ & 3 \\ \hline
$(QQQ)_4H\barH\barH$ & $(QQQ)_4^{\alpha\beta\gamma} H_\alpha \barH_\beta \barH_\gamma$ & & 1 \\ \hline
$(QQQ)_4LLLe$ & $(QQQ)_4^{\alpha\beta\gamma} L^m_\alpha L^n_\beta L^p_\gamma e^q$ & $m,n,p,q=1,2,3, n\le m,p \le n$ & 27 \\ \hline
$uudQdQd$ & $u^i_a u^j_b d^k_c Q^m_{f, \alpha}d^n_f Q^p_{g, \beta} d^q_g \epsilon^{abc} \eps$ & $\ba{l} i,j,k,m,n,p,q=1,2,3; \\ j<i, \mbox{antisymmetric}\{(m,n), (p,q)\} \ea$ & 324 \\ \hline
$(QQQ)_4LL\barH e$ & $(QQQ)_4^{\alpha\beta\gamma} L^m_\alpha L^n_\beta \barH_\gamma e^p$ & $m,n,p = 1,2,3, n \le m$ & 9 \\ \hline
$(QQQ)_4L\barH\barH e$ & $(QQQ)_4^{\alpha\beta\gamma} L^m_\alpha \barH_\beta \barH_\gamma e^n$ & $m,n=1,2,3$ &  9 \\ \hline
$(QQQ)_4\barH\barH\barH e$ & $(QQQ)_4^{\alpha\beta\gamma} \barH_\alpha \barH_\beta \barH_\gamma e^m$ & $m=1,2,3$ &  3 \\ \hline
\end{tabular}
\end{footnotesize}\end{center}}{\caption{\label{tbl:gio}
{\bf The set $D = \{r_i\}$ of generators of gauge invariant operators for the MSSM}.
The $28$ types are listed in the first column with the explicit sum and indexing given in the next two columns.
The final column counts the number of possible flavor combinations for each type.}}
\end{table}

We have attempted to implement the algorithm of Section~\ref{s:algo} on the MSSM directly.
The number of matter fields, the number of operators in the superpotential, and the size of the minimal set of gauge invariant operators all contribute to the complexity of the computational analysis.
Unfortunately, at present, these forces overwhelm the capacity of our desktop machines.
It should be emphasized that this is not a failure of the algorithm itself ---
it is simply a complicated case which will require more powerful computing resources to tackle.
This is work in progress.
For now, we make a first search for special structure within subsectors of the full theory.
This is a sensible course of action in any case even excluding the computational difficulties that we have mentioned.
Often in string models, for example, different sectors of low-energy effective theories come from different parts of the construction.
Any one of these sub-constructions may give rise to structure of the type we are searching for.

\subsection{One Generation}

A first simple case study one could undertake is to look at the situation where there is only one generation of particles, {\em i.e.}\ if all $i,j,k = 1$.
In this instance we may indeed use the simplified notation as in the left-most column of Table~\ref{tbl:gio} above without ambiguity.
Most of the gauge invariant operators listed anticommute to zero and the remaining list is
\beq \label{1gengio}
LH,\; H \barH, \; QdL, \; QuH, \; Qd\barH, \; L\barH e, \; QuQd, \; QuLe, \; Qu\barH e.
\eeq
We are now working in $R = \IC[\phi_1, \ldots, \phi_9]$, mapping to $S = \IC[r_1, \ldots, r_9]$, a much more manageable task indeed.
The renormalizable superpotential of~(\ref{renorm}) simplifies to
\beq\label{W-renorm-1gen}
W_{\rm minimal} = C^0 \sum_{\alpha, \beta} H_\alpha \barH_\beta \eps + C^1 \sum_{\alpha, \beta, a} Q_{a,\alpha} u^a H_\beta \eps +
                  C^2 \sum_{\alpha, \beta, a} Q_{a,\alpha} d_a \barH_\beta \eps + C^3 \sum_{\alpha, \beta} e L_{\alpha} \barH_\beta \eps.
\eeq

We may now freely input the above data into our algorithm.
For comparison, let us include some other terms into the superpotential in addition to those in~\eref{W-renorm-1gen}.
In particular there are five possible operators which are monomial in the gauge invariant operators of~\eref{1gengio}.
Adding any one of these to~\eref{W-renorm-1gen} represents a {\bf deformation} to the base superpotential.
We may consider~\eref{W-renorm-1gen} and these deformations as a {\bf class} of related theories.
We collect in Table~\ref{def-1gen} the perturbation considered to the base superpotential against the geometry this gives rise to in the supersymmetric vacuum space.

\begin{table}
{\begin{center}
\begin{tabular}{|c|c|c|}\hline
  $W_{\rm minimal}$ + ? & ${\rm dim}(\CM)$ & $\CM$ \\ \hline \hline
  0 & 1 & $\IC$ \\ \hline
  $LH$ & 0 & point \\ \hline
  $QdL$ & 0 & point \\ \hline
  $QuQd$ & 1 & $\IC$ \\ \hline
  $QuLe$ & 1 & $\IC$ \\ \hline
  $Qu\barH e$ & 1 & $\IC$ \\ \hline
\end{tabular}
\end{center}} {\caption{\label{def-1gen} {\bf Vacuum space geometry for one generation MSSM plus deformations}. }}
\end{table}

The above results do not seem too exciting.
In words, the result $\CM = \IC$ for the base case of~(\ref{W-renorm-1gen}) implies that there is a one parameter family of field vevs which are left undetermined in the supersymmetric vacuum space and these vevs determine the complex line.
Adding certain deformations, such as $LH$, to the superpotential reduces this space.
Specifically, the supersymmetric vacuum of $W_{\rm minimal} +LH$ is a point in a nine-dimensional complex space at which all nine fields in the theory must acquire a well-defined vev.

Perhaps a more interesting moduli space can be found by looking for structure within $W_{\rm minimal}$ by {\em reducing} the number of terms in the superpotential.
For example, if we neglect the Higgs mass term and consider
\beq W_{\rm minimal} = C^1 \sum_{\alpha, \beta, a} Q_{a,\alpha} u^a H_\beta \eps + C^2 \sum_{\alpha, \beta, a} Q_{a,\alpha} d_a \barH_\beta \eps +
                       C^3 \sum_{\alpha, \beta} e L_{\alpha} \barH_\beta \eps,
\eeq
then we find that ${\rm dim}(\CM) = 2$; in fact
\beq
\CM =\IC^2.
\eeq
As these cases are geometrically trivial we find that there is no interesting geometric structure in the supersymmetric vacuum space of a one generation minimal supersymmetric standard model with any reasonable phenomenology (matter Yukawa couplings).
This is potentially an encouraging result in its own right as it implies that any interesting structure which might be present in the full MSSM is crucially dependent on the existence of multiple families of Standard Model matter.
This gives us hope that we might be able to link the phenomenologically puzzling existence of non-trivial flavor structure to a higher dimensional explanation.

\subsection{Electroweak Sector}

Let us proceed to a less trivial example: the electroweak sector of the MSSM.\footnote{
The resulting theory is anomalous.
When fields are charged under an anomalous $U(1)$, the generalized Green--Schwarz mechanism~\cite{gs1,gs2,gsw} renders the theory consistent.
We have D-terms
$\sum_i \phi_i^\dagger \phi_i - \xi^2 = 0$.
The algorithm, as we have seen, implements D-flatness via a (symplectic) quotient and is unmodified by the addition of a Fayet--Iliopoulos term.
The analysis of the electroweak sector should be regarded as a survey of the vacuum space of a particular subsector of the full theory.
In any case, the electroweak theory is not the entirety of physics, and it is perfectly acceptable for a subsector to be anomalous.}
We set the fields $Q$, $u$, and $d$ to zero in Table~\ref{tbl:gio} and in the superpotential~\eref{renorm}.\footnote{
We note as an aside that when all the vevs of a field charged under a particular gauge group vanish identically, this gauge symmetry is preserved by the vacuum.
Thus setting the quark fields to zero is consistent with the unbroken $SU(3)_C$ symmetry of the Standard Model.
Insisting upon this feature of the vacuum space may be a useful {\em input} in the search for high-energy physics.
It should be remembered, however, that we are interested in the geometry of the supersymmetric rather than the non-supersymmetric vacuum.
As such, this restriction is not compulsory.
We thank David Berenstein for a discussion on this point.}
The fields are then $L$, $H$, $\barH$ and $e$ and the set $D$ of Table~\ref{tbl:gio} reduces to that of Table~\ref{gio-ew}.
There are $22$ holomorphic gauge invariant operators.
Thus we are studying $R = \IC[\phi_1, \ldots, \phi_{13}]$ mapping by the operators $D$ to an ideal of $S = \IC[r_1, \ldots, r_{22}]$.
The renormalizable superpotential is, from~\eref{renorm},
\beq\label{renorm-ew}
W_{\rm minimal} = C^0 \sum_{\alpha, \beta} H_\alpha \barH_\beta \eps + \sum_{i,j} C^3_{ij} e^i \sum_{\alpha, \beta} L^j_{\alpha} \barH_\beta \eps.
\eeq

\begin{table}
{\begin{center}
\begin{tabular}{|c||c|c|c|}\hline
\mbox{Type} & \mbox{Explicit Sum} & \mbox{Index} & \mbox{Number} \\
\hline \hline
$LH$  & $L^i_\alpha H_\beta \eps$ & $i=1,2,3$ & 3 \\ \hline
$H\barH$ & $H_\alpha \barH_\beta \eps$ & & 1  \\ \hline
$LLe$ & $L^i_\alpha L^j_\beta e^k \eps$ & $i,j=1,2,3;
k=1,\ldots,j-1$ & 9  \\ \hline
$L\barH e$ & $L^i_\alpha \barH_\beta \eps e^j$ & $i,j=1,2,3$ & 9 \\
\hline
\end{tabular}
\end{center}}{\caption{\label{gio-ew}{\bf The set $D = \{r_i\}$ of generators of gauge invariant operators for the electroweak sector of the MSSM}.}}
\end{table}

We find that the resulting moduli space $\CM$, as an affine variety, is five-dimensional.
It is in fact an affine cone over a base manifold $\CB$ of dimension four.
As a projective variety, $\CB$ has degree six and is described by the (non-complete) intersection of six quadratics in $\IP^8$.
That is to say,
\begin{equation}
\CM_{\rm EW} = (8|5,6|2^6). \label{M_EW}
\end{equation}
Because of the high dimensionality of $\CM$ in this case, the moduli space is
mathematically difficult to study.
The equations for the base manifold do not indicate obvious special structure.

As in the one generation case, we could add various deformations to the superpotential from the list of gauge invariant operators in Table~\ref{gio-ew}.
In particular there are two additional types of operators which we could add to the superpotential~\eref{renorm-ew}, $LH$ and $LLe$.
Both types of operators $LH$ and $LLe$ violate R-parity, which in this simple theory is equivalent to lepton number.
The principal supersymmetric candidate for cold dark matter is lost with their inclusion, but neither term (alone or in conjunction) generates proton decay.
Furthermore, the dimensionless coefficients of these terms are only weakly and indirectly constrained by observation.
Both can contribute new supersymmetric contributions to $\mu$ - $e$ conversion, rare leptonic decays, leptonic branching fractions for mesons, and the anomalous magnetic moment of the muon.
But these constraints (for $100$ GeV mass scales) are no more than $O(10^{-3})$ for the individual dimensionless couplings~\cite{Barbier:2004ez}, so there is no {\em a priori} reason to forbid them from the point of view of phenomenology.

For each of three cases, we repeat the analysis performed above and tabulate the results in Table~\ref{def-ew}.
Once again we verify our intuition that adding additional interactions to the superpotential reduces the dimensionality of the resulting vacuum manifold.
In this case the effect is dramatic.
The terms $LH$ and $LLe$ reduce a rich structure to a trivial one.

\begin{table}
{\begin{center}
\begin{tabular}{|c|c|c|}\hline
  $W_{\rm minimal}$ + ? & ${\rm dim}(\CM)$ & $\CM$ \\ \hline \hline
  0 & 5 & $(8|5,6|2^6)$ \\ \hline
  $LH$ & 1 & $\IC$ \\ \hline
  $LLe$ & 0 & point \\ \hline
  $LH$ + $LLe$ & 0 & point \\ \hline
\end{tabular}
\end{center}} {\caption{\label{def-ew} {\bf Vacuum space geometry for renormalizable electroweak sector of the MSSM plus deformations}. }}
\end{table}

\subsubsection{Lifting the Higgs Terms}\label{s:EW-ord4}

Let us consider another class of deformations of the base case in~(\ref{renorm-ew}).
By adding the allowed order four terms from the list in Table~\ref{gio-ew}, flat directions associated with field vevs for the Higgs multiplets can be lifted, meaning that $\langle H \rangle$ and $\langle \barH \rangle$ are constrained to vanish.\footnote{
In principle, the $\mu$-term in the renormalizable superpotential lifts the Higgs by itself.
However, because $\mu$ is of order the electroweak scale, the term produces a negligible contribution to the scalar potential.}
Let us therefore take the superpotential
\beq
W = W_{\rm minimal} + \lambda(H_\alpha \barH_\beta \eps)^2 + \lambda^{ij} (L_i H_\alpha) (L_j H_\beta) \eps, \label{lift} \eeq
with $\lambda$ and $\lambda^{ij}$ flavor mixing coefficients.
This is the most general superpotential possible at this order which is consistent with R-parity.
The new terms in \eref{lift} are natural to consider precisely because both arise in well-motivated contexts in which heavy Standard Model singlets are integrated out of the theory:
a singlet that generates the $\mu$-term as in the NMSSM for the first term or (famously) a right-handed neutrino to generate the canonical see-saw mechanism for the second term.
We expect that the addition of these new interaction terms to the superpotential will generically make the F-flatness conditions more involved and thus harder to satisfy.
This should na\"{\i}vely reduce the dimensionality of the resulting vacuum space.
Indeed, in this case we find that the moduli space of the theory defined by~(\ref{lift}) is three-dimensional.
More precisely, $\CM$ is an affine cone over a base surface $\CB$.
The geometry of $\CB$ is as follows.
It is given by the (non-complete) intersection of six quadratics in $\IP^5$.
The degree of $\CB$ is four.
Thus,
\beq
\CM_{\rm lifted \; EW} = (5|3,4|2^6). \label{M_lift}
\eeq

Since $\CB$ is a (compact) projective variety we can readily compute its Hodge diamond
\begin{equation}
h^{p,q}(\CB) \quad = \quad
{\begin{array}{ccccc}
&&h^{0,0}&& \\
&h^{0,1}&&h^{0,1}& \\
h^{0,2}&&h^{1,1}&&h^{0,2} \\
&h^{0,1}&&h^{0,1}& \\
&&h^{0,0}&& \\
\end{array}}
\quad = \quad
{\begin{array}{ccccc}
&&1&& \\
&0&&0& \\
0&&1&&0 \\
&0&&0& \\
&&1&& \\
\end{array}}.
\label{hodge}
\end{equation}
According to the classification of degree $d$ surfaces in $\IP^{d+1}$~\cite{hart}, there are only three possible degree four surfaces in $\IP^5$.
These are
\label{list}
\begin{enumerate}
\item The {\em Veronese surface}, which is an embedding of $\IP^2$ in $\IP^5$:
\be
\ba{ccc}
\IP^{2} & \rightarrow & \IP^{5} \cr
[x_0:x_1:x_2] & \mapsto &
[{x_0}^2 : x_0x_1 : {x_1}^2 : x_0x_2 : x_1x_2 : {x_2}^2];\label{veronese}
\ea
\ee
\item  The rational scroll $S(1,3)$, formed by lines joining a point to the twisted cubic curve;
\item The rational scroll $S(2,2)$, formed from two conics in $\IP^5$.  \end{enumerate}
All three candidates have the same Hodge diamond as above and all have Hilbert polynomial $-3H(\IP^1) + 4H(\IP^2)$.
Moreover, all crude birational invariants for the three surfaces are also the same.
So how do we distinguish amongst them?
It turns out that we can do so by studying the Fano variety of lines\footnote{
The {\em Fano variety of lines} $F(\CB)$ is just the variety parameterizing lines on $\CB$.
This should not be confused with a Fano variety with ample anticanonical bundle.}
associated with these three surfaces~\cite{mac}.
In the case of the manifold defined by~(\ref{M_lift}), the Hilbert polynomial of the Fano variety of lines for $\CB$ vanishes.
This is true only of the first case of the three possibilities; therefore we conclude that $\CB$ is the Veronese surface.
Thus we have been able not only to compute the dimensionality of the moduli space of the MSSM electroweak sector
(which could have been done by traditional methods such as counting F-terms),
but we have also extracted the full information about this space and conclusively identified what it is.
Because the Veronese surface is an embedding of $\IP^2$ into $\IP^5$, it inherits the Hodge structure of $\IP^2$; however it is algebraically and geometrically more complicated as a manifold.

As an aside we note that since the Veronese surface is defined by a degree two embedding of $\IP^2$ into $\IP^5$, the affine cone $\CM$ over $\CB$ is simply
\beq
X = \cO_{\IP^2}(-2).
\eeq
We remark, that the line bundle $\CO_{\IP^2}(-3)$, which happens to be the canonical bundle of $\IP^2$, is a familiar object in string theory.
It is a Calabi--Yau resolution of the orbifold $\IC^3/\IZ_3$ which has been extensively studied in the context of D-brane probes \cite{dgm}.
This case was examined in detail in Section~\ref{s:z3}.
On the other hand, our present cone over the Veronese surface, $\CO_{\IP^2}(-2)$, is \emph{not} a Calabi--Yau space.
As such constructing the electroweak model under consideration in the Coulomb phase of a single D3-brane on a singularity would be strictly impossible.

\subsubsection{Adding the Right-Handed Neutrino}
\label{sec:rhv}

Another general class of deformation one could consider is to add new degrees of freedom in terms of chiral superfields.
For example, in the electroweak sector we may wish to include right-handed neutrinos.
Though these fields are singlets of the Standard Model gauge group, models of neutrino mass typically involve superpotential couplings of these fields with other fields in the electroweak sector.
Thus we will add three generations of right-handed neutrino $\nu^{i=1,2,3}$ to the gauge invariant operators listed in Table~\ref{gio-ew}.

For our baseline superpotential we will adopt the couplings required for the standard see-saw mechanism~\cite{GRS,Yanagida,Valle1,Valle2}.
The analogue of~\eref{renorm-ew} then becomes
\beq\label{renorm-ew-nu}
W_{\rm minimal} = C^0 \sum_{\alpha, \beta} H^\alpha \barH^\beta \eps + \sum_{i,j} C^3_{ij} e^i \sum_{\alpha, \beta} L^j_{\alpha} \barH_\beta \eps +
                  \sum_{i,j} C^4_{ij} \nu^i \nu^j + \sum_i C^5_{ij} \nu^i \sum_{\alpha, \beta} L^j_\alpha H_\beta \eps.
\eeq
We find that with the addition of these extra degrees of freedom the moduli space changes dramatically at the renormalizable level.
It is now three-dimensional, is the intersection of six quadratics in $\IP^5$, and is degree four.
In fact, it is our familiar cone over the Veronese surface.
Therefore, with the addition of the right-handed neutrino, the renormalizable superpotential, without order four terms, gives the same three-dimensional variety which we found previously.

\begin{table}
{\begin{center}
\begin{tabular}{|c|c|c|}\hline
  $W_{\rm minimal}$ + ? & ${\rm dim}(\CM)$ & $\CM$ \\ \hline \hline
  0 & 3 & $(5|3,4|2^6)$ \\ \hline
  $LH$ & 2 & $(2|2,2|2) \simeq \{ \mbox{cone over }
  \IP^1 \}$ \\ \hline
  $LLe$ & 0 & point \\ \hline
  $H \barH \nu$ & 3 & $(5|3,4|2^{6})$ \\ \hline
 $LH +H \barH \nu$ & 2 &  $(2|2,2|2) \simeq \{ \mbox{cone over }
  \IP^1 \}$ \\ \hline
\end{tabular}
\end{center}} {\caption{\label{def-nu} {\bf Vacuum space geometry for renormalizable electroweak sector of the MSSM with right-handed neutrinos, plus deformations}. }}
\end{table}

Let us now consider how robust the Veronese surface is to further deformations of the superpotential away from~(\ref{renorm-ew-nu}).
For example, we may consider adding renormalizable terms analogous to those considered in Table~\ref{def-ew}.
The results of this analysis are compiled in Table~\ref{def-nu}.
Some properties of this table deserve mention.
First it is again the case that inclusion of R-parity violating operators $LLe$ and $LH$ reduces the vacuum space to a trivial one.
However, in this example there is the possibility of adding the operator $H\barH\nu$ without altering the vacuum space at all.
This is an example of an {\bf ineffectual} ``flat'' direction in the space of all possible deformations to the base theory~(\ref{renorm-ew-nu}).
Under an ineffectual deformation, although the superpotential and interactions of the theory change, the vacuum moduli space remains the same.
Note that this implies that the special structure in this example ---
that is, the fact that the vacuum space is simply described by the cone over the Veronese embedding of $\IP^2$ in $\IP^5$ ---
is {\em not} merely a reflection of some underlying gauge symmetry such as $SO(10)$ since this operator would be forbidden in the case where $\nu$ arises from the $\mathbf{16}$ of $SO(10)$ but the Higgs doublets are associated with $\mathbf{10}$ representations.

In fact, this particular example illustrates the possible power in associating special geometric structure in the vacuum space of a theory with low-energy phenomenology.
First, we know now that neutrinos have mass~\cite{neutrino}, and the theory considered in~\eref{renorm-ew-nu} is one of the leading candidates for a phenomenological description of this fact.
Second, the special structure which has appeared in this case emerges at the renormalizable order.
This is important due to the nature of the globally supersymmetric limit we are considering of these theories.
Since we expect Nature to be described by supergravity at energies well below the string scale, there are various sources of $M_{\rm Pl}$ suppressed corrections to the moduli space geometry which we are not considering here.
In this sense including the order four terms in Section~\ref{s:EW-ord4} could be viewed as somewhat problematic in that there are other corrections of the same size which are being neglected.\footnote{
We point out, however, that in adding to the superpotential \eref{renorm-ew-nu} at mass level four either or both of the R-parity preserving terms from before, $(H\barH)^2$ or $(LH)^2$, the Veronese surface persists as the vacuum manifold.
These are ineffectual terms for geometry.}
This criticism does not apply to the example of the field theory with right-handed neutrino that we have presented in this subsection.

\subsection{A Systematic Analysis of the Electroweak MSSM}

We have seen indications that the geometry of the supersymmetric vacuum space of the electroweak MSSM is special in at least some sense.
The minimal superpotential~\eref{renorm-ew} with $\mu$-term and $L\barH e$ Yukawa terms yields the vacuum manifold $\CM_{EW} = (8|5, 6|2^6)$.
Only when we incorporate into the superpotential either (a) R-parity preserving terms at mass level four, or (b) Majorana and Dirac mass couplings involving right-handed neutrinos do we obtain as the vacuum manifold the affine cone over the Veronese surface, which is one of the simplest non-trivial structures encountered in algebraic geometry.
Each of the deformations to~\eref{renorm-ew} that lead to the Veronese surface that we have so far examined are motivated by phenomenological considerations:
in the first case, R-parity ensures stability of the proton and preserves a stable candidate for cold dark matter;
in the second, neutrinos acquire mass through a canonical see-saw.
Interactions demanded by particle phenomenology therefore refine the geometry of the vacuum manifold from the complicated $(8|5, 6|2^6)$ variety to the simpler Veronese surface, whose topological invariants are precisely those of $\IP^2$.
The Hodge diamond~\eref{hodge} consists only of $0$s and $1$s.
This geometry is {\em special} in that there is no simpler two-dimensional complex variety (other than the trivial $\IP^2$). 
Deforming the physical theory by adding additional operators that are not present in the minimal models in general destroys the geometric structure we have uncovered and renders the moduli space of the theory trivial.
The additional operators that we include do not respect lepton number and are thus disfavored from the point of view of phenomenology (or at least heavily suppressed).

We must ask ourselves whether the coincidence of geometry and phenomenology in this example is accidental or consequential.
In other words, {\em how special is the Veronese surface?}

We will answer this question by studying a class of theories with the {\em same} field content, but {\em different} interactions among the fields.
There are $16$ matter degrees of freedom in total: the Higgs doublets $H_\alpha$ and $\barH_\alpha$ and the leptons $L^i_\alpha$, $e^i$, and $\nu^i$.
For the interactions, we will consider possible superpotentials consistent with gauge invariance up to mass level three (renormalizable order).
In order to restrict the cases we must analyze, we impose the extra condition of {\bf flavor democracy}, which means that if an operator with a given flavor structure is present in the superpotential, then that interaction is present for all possible values of the flavor indices.\footnote{
We do not consider textures.
For algebraic geometry, what is important is whether an operator is present or absent in the superpotential.
The relative weight of its coefficient with respect to the other coefficients will not matter for the analysis of topological features of the manifold.}
This is simply to say that we do not write down a Yukawa term for the electron without also writing down Yukawa terms for the muon and tau.
Each term and each assignment of flavor indices appear in the superpotential with a random coefficient.
As before, for the purposes of determining the geometry, the mass dimension of the coefficient will not matter.

Working to renormalizable order, we can write superpotentials involving nine types of operators.
These are listed in Table~\ref{tbl:512}.
Each of these are either present or absent in the superpotential, so, subject to flavor democracy, there are $2^9 = 512$ cases to explore.
Most of the superpotentials are of course completely unphysical.
We want to test whether interesting geometry correlates to interesting physics, so of course we should study a class of theories involving both superpotentials that might make sense as a description of Nature and those that manifestly will not.

\begin{table}[thb]
\begin{center}
\begin{tabular}{|c|c|}\hline
Mass Level & Interactions \\ \hline \hline
$1$ & $\nu$ \\ \hline
$2$ & $L H$, $H \barH$, $\nu^2$ \\ \hline
$3$ & $L \barH e$, $L L e$, $L H \nu$, $H \barH \nu$, $\nu^3$ \\ \hline
\end{tabular}
\end{center}
{\caption{\label{tbl:512} {\bf Renormalizable interactions in the electroweak sector.}}}
\end{table}

Of the $512$ superpotentials we can write from operators in Table~\ref{tbl:512}, we have calculated the dimension of the supersymmetric vacuum space in $445$ cases ($87$\%).
The reason we have not calculated the dimension for the remaining $67$ is simply that the Gr\"obner basis calculations are more intensive in these instances.
Experience has taught us that such cases tend to give varieties of very low dimension, as might be expected as more equations admit a more restrictive solution space.
As such, completing the analysis of all $512$ cases would require a greater commitment of time and computer resources while furnishing little new information.
We have therefore restricted ourselves to the $445$ cases that are more readily analyzed.

In order to compare the geometry of the moduli space of these theories to those that we have examined in Section~\ref{sec:rhv}, we will focus our attention on the three-dimensional manifolds.
Of the $58$ three-dimensional varieties in the data set, we have performed a detailed geometric analysis of $46$.
Most of the vacuum manifolds are flat space or multiple copies of flat space, but among the geometries are also a number of conifolds.
As well, we find examples of the Veronese surface that we tabulate in Table~\ref{tbl:ver}.\footnote{
In general, the vacuum spaces are reducible varieties, which is to say that they are the union of separate irreducible varieties.
To properly analyze the geometry, one must first split the reducible variety into its irreducible components using a technique known as {\em primary decomposition} (see~\cite{cox} for details) and then analyze each component separately.
When we say that $\CM$ is an affine cone over Veronese $\cup$ $\BC$, what is meant is that one component of the base is the Veronese surface and another is the complex line.}
Interestingly, the rational scrolls never appear
(these are possibilities~$2$ and $3$ from the list of degree four surfaces in $\IP^5$ from page~\pageref{list}).

\begin{table}[thb]
\begin{center}
\begin{tabular}{|c|c|c|}\hline
Case & Superpotential terms & %
$\CB$ \\ \hline \hline
$1$ & $H \barH$, $\nu^2$, $L \barH e$, $L H \nu$ & Veronese \\ \hline %
$2$ & $H \barH$, $\nu^2$, $L \barH e$, $L H \nu$, $H \barH \nu$ & Veronese \\ \hline %
$3$ & $H \barH$, $L \barH e$, $L H \nu$, $\nu^3$ & Veronese \\ \hline %
$4$ & $\nu^2$, $L \barH e$, $L H \nu$ & Veronese $\cup$ $\BC$ \\ \hline %
$5$ & $\nu^2$, $L \barH e$, $L H \nu$, $H \barH \nu$ & Veronese $\cup$ $\BC$ \\ \hline %
$6$ & $L \barH e$, $L H \nu$, $\nu^3$ & Veronese $\cup$ $\BC$ \\ \hline %
$7$ & $L \barH e$, $L H \nu$, $H \barH \nu$, $\nu^3$ & Veronese $\cup$ $\BC$ \\ \hline %
$8$ & $\nu^2$, $L L e$, $L H \nu$, $H \barH \nu$ & Veronese \\ \hline %
\end{tabular}
\end{center}
{\caption{\label{tbl:ver}{\bf Veronese surfaces in the electroweak sector.
The vacuum moduli space $\CM$ is an affine cone over $\CB$.}}}
\end{table}

In this list, Case~$1$ is the MSSM with a right-handed neutrino sector.
Case~$2$ includes the ineffectual lepton number violating $H\barH\nu$ deformation that we encountered in Section~\ref{sec:rhv}.
Case~$3$ replaces the $\nu^2$ Majorana term with a cubic neutrino interaction, which also clearly fails to conserve lepton number.
Case~$4$ drops the $\mu$-term.
This changes the vacuum geometry from the Veronese surface to the Veronese surface plus a complex line.
Case~$5$ drops the $\mu$-term and adds the $H\barH\nu$ deformation, which is again ineffectual.
Case~$6$ drops the $\mu$-term and the $\nu^2$ Majorana term and includes the $\nu^3$ vertex.
Case~$7$ is an ineffectual $H\barH\nu$ deformation to Case 6.
Finally, Case~$8$ is completely unphysical as it does not even include Yukawa couplings for electrons, muons, and taus.
While adding ineffectual deformations restricts to a particular class of theories, this list is certainly more general.
At a minimum, it involves several classes.

What should we conclude from this exercise?
We will first make some comments about the top-down approach to model building.
Having the Veronese surface as the vacuum variety does not uniquely determine the interactions in the theory.
With identical field content, different superpotentials give rise to the same moduli space of F- and D-flat configurations.
This is perhaps not a surprise since we know that the same quiver diagram (field content of the gauge theory) associates to different superpotentials based on the structure of the underlying singularity~\cite{bjl0}.
Nevertheless, the catalogue of superpotentials and geometries constrains the options for model building.
If, for example, one wanted to consider a theory such that the vacuum space was the conifold, only certain superpotentials possess this property at renormalizable order.
The conifold examples that we have found share the feature that the $L\barH e$ Yukawa coupling does not appear in the superpotential.
Because these examples do not exhibit any reasonable phenomenology in the low-energy effective theory, we do not present these data here or attempt a top-down string construction.

If we were to place D-branes at singularities, we may use the worldvolume gauge theory on the brane as a probe of higher dimensional physics.
In these models, the equation for the vacuum space as an algebraic variety is the same as the equation for the singularity whose transverse directions the D-brane probes.
The gauge theory captures motion on the resolved space \cite{DM}.
Though one comes close, we know that the MSSM does not arise as a quiver gauge theory corresponding to a D-brane at an isolated orbifold singularity on a Calabi--Yau threefold~\cite{BJL}.
{\em The MSSM electroweak sector with right-handed neutrinos does not arise in this way either, for the affine cone over the Veronese surface is not Calabi--Yau.}
We conclude as a result of these investigations that a method of top-down model building that results in a three-dimensional Calabi--Yau vacuum variety cannot be used to construct low-energy effective electroweak theory.
These observations supply further evidence for believing that a top-down construction of the MSSM will demand new advances in model building.

There are lessons for the bottom-up approach here as well.
The philosophy we have adopted is that {\em if} we have a vacuum geometry with suitably special features that are unexplained by discrete and continuous symmetries in field theory, {\em then} we should regard this as an imprint of high-energy physics in the low-energy theory and use this to make predictions about other low-energy phenomenology.
We posit that the operators that preserve the structure --- namely the ineffectual ones --- give new interactions in the theory while other equally gauge invariant operators do not.
In the electroweak sector, we cannot say that this methodology should be applied for the simple reason that a single class of deformations to the superpotential~\eref{renorm-ew-nu} does not associate uniquely to the Veronese surface.
In other words, the vacuum geometry, though simple and geometrically appealing, is not suitably special enough for us to pursue a dedicated study of its deformation theory and promote the Veronese geometry to a principle regarding the ultraviolet completion of the gauge theory.
We are thus unable to test the hypothesis we have made within the electroweak sector.

Here, we have considered a theory with a small number of fields ($16$ of them) and a small minimal basis of holomorphic gauge invariant operators ($25$ of these) as inputs.
It may be that in such a theory the possibilities for three-dimensional vacuum spaces are fairly limited, so perhaps it is not a surprise that the Veronese surface appears many times rather than just once.
A more complicated theory, or a more complicated sector of the MSSM, may provide a better arena to explore the utility of vacuum geometry for phenomenology.\footnote{
We thank Nima Arkani-Hamed for comments on this point.}
\section{Discussion}

Our goals in studying the supersymmetric vacuum space of phenomenological theories are threefold.
First, we wish to search for special geometric structures in the vacuum moduli space of $\CN=1$ gauge theories.
Such structures are common in the low-energy effective field theories descending from string and M-theory, but unmotivated in field theories in general.
As such this structure, if present, would constitute significant evidence for string physics underlying the Standard Model.

Should such structures be found, we can then move on to our second goal.
{\em If} the geometry is suitably {\em special}, meaning that it is unlikely to be a result of chance, {\em then} we might conclude that this geometry is a fundamental consequence of the high-energy completion that Nature has chosen.
Therefore, when considering higher order terms in the superpotential and K\"ahler potential, the theory necessarily restricts to operators that respect this structure.
The slogan is no longer that anything not forbidden by gauge invariance is allowed.
Rather, the standard we advocate is that {\em anything not forbidden by gauge invariance and consistent with vacuum geometry is allowed.}
Application of this principle would then predict the absence of certain operators in the theory --- and thus the suppression of the effects they mediate.

$\CN=1$ vacuum manifolds as well often enjoy global isometries, continuous or discrete, that are not {\em a priori} evident from the superpotential.
For example, in D-brane probe scenarios, one could find such symmetries of the vacuum variety, and then rearrange or redefine fields in the superpotential to exhibit the hidden global symmetries explicitly~\cite{fhk}.
The techniques we have presented may also facilitate the search for manifestly symmetric forms of the Lagrangian, but these symmetries are logically distinct from the use and utility of geometry as a selection principle in its own right.
Indeed, by considering discrete symmetries in this language our approach makes it easier for string model-builders to engineer their presence in the low-energy effective Lagrangian.

Finally, if no special structure is found, we can still make powerful statements.
Because many string constructions are known to produce theories whose vacuum spaces exhibit special structure it is clear that the issue of what sorts of low-energy theories are possible and which are impossible in a given construction is illuminated by a search such as the one we have initiated.
Developing a catalogue of theories and their vacuum geometries is the third goal of this research.
This is in the spirit of~\cite{swamp} in isolating the physical theories that have an embedding within a string framework.

In this paper we have presented in detail our method for computing the supersymmetric vacuum space of $\CN=1$ gauge theories.
This discussion reduces the analysis to a pure algorithmic process involving Gr\"obner basis techniques.
We have applied this method to various sectors of the MSSM, providing the first new results in this regard since the dimension was calculated in 1995~\cite{mssm}.
In particular, we have analyzed the geometry of the vacuum space of the full one-generation MSSM and, in some detail, of electroweak theory, with and without right-handed neutrinos.
These examples were presented as illustrations of how our algorithms can be implemented and as an initial scan for new physics in the vacuum space.
While interesting exercises in their own right, these examples are not intended to be the final expression of our goals.
Nevertheless, several interesting lessons can already be identified.

The first lesson is that special structure of any kind is indeed rare.
The $SU(2)_L \times U(1)_Y$ sector of the MSSM is a relatively simple system.
Most superpotentials that can be constructed from the fields of this system yield a trivial vacuum space.
Of the $512$ (renormalizable) superpotentials, we have found that only eight of these theories have a vacuum space which is a cone over the Veronese base.
That these eight possibilities include the minimal electroweak model of the MSSM with the canonical see-saw mechanism is encouraging, as is the fact that higher-order terms which preserve the special structure can be identified.
From the point of view of string model-building, it is also interesting that this is a surface of three complex dimensions.
Of course this is merely a subsector of a truly realistic model, such as the full three-generation MSSM.
Yet within this toy-model context would we be wise to consider the Veronese structure as somehow fundamental and use its presence as a guide to further model-building?

The answer, unfortunately, is profoundly unclear at this point.
The presence of the Veronese structure does not specify one superpotential and a class of deformations to this structure uniquely.
We see, however, that this structure is not identical to gauge invariance, since the other $504$ superpotentials also enjoyed the $SU(2)_L \times U(1)_Y$ invariance.
The Veronese structure does not arise from including {\em all} possible gauge invariant terms to this mass dimension, but only certain subsets.
These subsets are not distinguished by any obvious discrete symmetry properties, nor by how they could be embedded in a larger gauge theory such as $SO(10)$.
Indeed, presence or absence of special geometrical structure is clearly related, but not identical to, these properties of the theory.
Even more interesting is the relation between the presence of multiple generations and vacuum space geometry:
any special structure which may exist in the Standard Model must be intrinsically linked to the presence of multiple generations and the accompanying flavor degrees of freedom.
This link between flavor physics and geometry gives us hope that a new perspective on this difficult subject may evolve from this work.

With these examples, we certainly conclude that special structure is indeed rare, but it remains unclear how this correlates with other phenomenological properties of the theory.
More investigation into wider classes of models may help settle the issue and this is work currently in progress.
A more general survey of the MSSM will require an improvement in the computational implementation of our algorithm due to the exponential growth of the Gr\"obner basis complexity as the number of gauge invariant operators and F-term constraints increase.
Nevertheless, many laboratories are accessible:
GUTs with unified representations,
neutrino mass models,
pure supersymmetric QCD,
and the MSSM extended by additional $U(1)$ symmetries (which reduce the number of gauge invariant operators and allowed superpotential couplings).
In the meantime, the power of our approach to achieve the third of our goals is already apparent and conclusive.
In this example it is already possible to state unequivocally that the electroweak MSSM does not arise from a D3-brane on a local singularity on a Calabi--Yau.

It is our expectation that some form of string physics underlies the four-dimensional effective field theory at low-energies.
Therefore we anticipate that the enterprise of building string models that generate theories such as the MSSM (or whatever supersymmetric model may be revealed at the LHC) can only be accelerated by an investigation of the vacuum manifold of those theories in the manner we propose here.
We also expect that whatever effective theory is eventually the theory of physics beyond the Standard Model, it will likely exhibit special geometric structure.
This is simply a statement that the superpotential, gauge group, and matter representations of the theory are ultimately determined by the (non-trivial) geometry of some compact space transverse to our $(3+1)$-dimensional world.
It is our hope that these structures can be determined to correlate with certain phenomenological properties of the resulting effective Lagrangians, thereby elucidating a new pathway for connecting string constructions to the observable world.

\section*{Acknowledgments}
We thank
Nima Arkani-Hamed,
David Berenstein,
Philip Candelas,
Xenia de la Ossa,
Nick Dorey,
Steve Martin,
Fernando Quevedo,
Miles Reid, and
Graham Ross
for helpful discussions.
J.G.~is supported by CNRS.
Y.H.H.~is supported by the FitzJames Fellowship at Merton College, Oxford.
V.J.~is supported by PPARC.
B.D.N.~is supported by the U.S.~Department of Energy under Grant No.~DOE-EY-76-02-3071.

\appendix
\section{A Brief Guide to Algebraic Geometry}

As our investigation is heavily interdisciplinary, the linguistic barrier may necessitate a bit of activation energy.
The greatest hurdle is perhaps the terminology and nomenclature of (computational) algebraic geometry.
This is especially likely to be true for the phenomenologist.
It is therefore expedient to gather together some facts in an instructive and non-rigorous manner and produce a skeleton key to the mathematics invoked.
We emphasize that the concepts are straightforward:
at the core this is only polynomial arithmetic.

\subsection{Affine Varieties and Polynomial Rings}
The key idea behind algebraic geometry is to describe a space (for example  a manifold) in terms of the vanishing of some polynomials.
For example let us start by describing the concept of an {\bf affine variety}.
An affine variety is a set of points in $n$-dimensional complex space $\IC^n$ with coordinates $(x_1,\ldots,x_n)$, which satisfy a set of polynomial equations of the form $f_i(x_1,\ldots,x_n)=0$.
An affine variety is thus loosely speaking a submanifold of flat space which is defined as the locus on which a set of polynomials vanish.
It can be shown that the number of these polynomial equations can always be taken to be finite to describe any particular submanifold ${\cal M}\subseteq \IC^n$.

The purpose of algebraic geometry is to study geometrical properties of the variety $\CM$, in the language of commutative algebra.
This is because there is another way of describing such a submanifold of flat space which turns out to be more powerful when analyzing properties of the space.
Instead of defining $\CM$ in terms of the vanishing of a finite set of polynomials, we specify it by the set of {\it all} polynomials which vanish on $\CM$.
The resulting set of polynomials, which we shall refer to as $I({\cal M})$, is an example of a {\bf polynomial ring}.

We recall that a {\bf ring} is simply a set with addition and multiplication.
The set of polynomials is clearly a ring since adding and multiplying polynomials together return a polynomial.
We denote the ring of polynomials in $n$ variables with coefficients in $\IC$ as $\IC[x_1,\ldots,x_n]$.
Now, $I({\cal M})$ is clearly a subset of $\IC[x_1,\ldots,x_n]$.
In fact, $I({\cal M})$ is an {\bf ideal} of $\IC[x_1,\ldots,x_n]$.
This simply means that $I({\cal M})$ is a closed subset in the following sense:
multiplying any element of $I({\cal M})$ by an element of $\IC[x_1,\ldots,x_n]$ remains in $I({\cal M})$.

Much like normal subgroups, because $I({\cal M})$ is an ideal of $\IC[x_1,\ldots,x_n]$, it is possible to define the {\bf quotient ring} $\IC[x_1,\ldots,x_n]/I({\cal M})$.
This is defined to be the ring of all polynomials in the variables $x_1,\ldots,x_n$ where two polynomials are considered equivalent if they differ by a member of the ideal $I({\cal M})$.
It is not difficult to verify that this resulting object is also a polynomial ring.

The final object, the quotient ring $\IC[x_1,\ldots,x_n]/I({\cal M})$, encodes all of the geometrical information about ${\cal M}$ in an algebraically powerful package.
For example, the ideals of this quotient ring are in one-to-one correspondence, in a similar manner to the correspondence between ${\cal M}$ and  $I({\cal M})$ described above, with submanifolds of ${\cal M}$.
The most important aspect of this correspondence is the following.
The smallest possible submanifolds, the points of $\CM$, correspond to {\bf maximal ideals} of the quotient ring $\IC[x_1,\ldots,x_n]/I({\cal M})$.
A maximal ideal is not contained in any other ideal, except for the trivial one, namely the ring itself.
A maximal ideal is also a prime ideal, which for complex algebraic geometry means that the ideal corresponds to an {\bf irreducible variety}, one that cannot be decomposed as the union of other non-empty algebraic varieties.

In a polynomial ring, all ideals are {\bf finitely generated}.
To introduce some notation, we denote an ideal which is generated by the elements $f_1, \dots, f_n$ by $F = \langle \langle f_1, \dots, f_n \rangle \rangle$.
In other words this is the ideal which vanishes on the affine variety defined by $f_1 = \dots = f_n = 0$.
We see that this formalism first of all naturally adapts to the concept of F-terms.
The vanishing of $n$ F-terms defines the ideal $F$, and the variety of F-flatness corresponds to the quotient ring $\Fflat = \IC[x_1,\ldots,x_n] / F$.

\subsection{Projective Varieties and Affine Cones}

While easy to define, affine varieties are sometimes difficult to work with in calculations.
This is ultimately because non-compact embedding spaces can lead to various difficulties, such as points escaping off to infinity.
One of the advantages of using {\bf projective varieties} is to remove such difficulties by ``compactifying'' the varieties.

We recall that {\bf projective space} $\IP^n$ is the space of one-dimensional complex vector subspaces of $\IC^{n+1}$.
Points in projective space are labeled by coordinates $[x_0 : x_1 : \ldots : x_n]$,
and $[x_0 : x_1 : \ldots : x_n] = [y_0 : y_1 : \ldots : y_n]$ when there is a non-zero $\lambda\in\IC$ such that $y_i = \lambda x_i$ for all $i$.
The equivalence classes of points define {\bf homogeneous coordinates}.

Projective varieties are defined in a manner similar to their affine counterparts, but slightly more care is required in the definition.
One cannot simply talk about polynomial equations on projective space.
If a set of polynomial equations in the homogeneous variables is {\bf homogeneous}, which is to say that all the monomials have the same total degree, then we can talk instead about the loci of zeros of such equations.
This locus is invariant under the rescaling of the homogeneous coordinates.
Thus we can define a projective variety as a set of points in projective space where a series of homogeneous polynomials in the projective coordinates vanishes.
All of the objects described above in connection with the affine variety ${\cal M}$ then have analogues in the case of a projective variety.

Since we are computing the vacuum moduli space embedded into the space of fields, which is clearly $\IC^n$, we are interested in affine varieties.
Why then have we introduced projective varieties here?
Our goal is to introduce the concept of an {\bf affine cone}, an object we make extensive use of in the paper.
The reason for adopting this setup is that projective algebraic geometry, due to compactness, is much easier to handle, both conceptually and using computer packages.
We are computing the local properties of the vacuum space.
It will often (but not always!)\ be the case that the moduli space is an an affine cone over some compact projective space whose properties we can directly calculate.
Hence we have introduced the notation in \eref{eq:ac}.

The base of an affine cone is simply defined by considering an affine variety and then taking the variables $x_1,\ldots,x_n$ to be homogeneous coordinates on projective space.
From the comments on projective varieties above we see that this procedure is only possible if the equations defining the original affine variety are homogeneous.
The radial direction of the cone can then be thought of as the scale of the projective space (with the point at zero scale put back in).
As a simple example consider the case where we have two variables $x_1$ and $x_2$ and where the defining equations of the original affine variety are trivial.
The base of the resulting affine cone will then simply be $\IP^1$, with projective coordinates $[x_1: x_2]$.
The space $\IC^2$, with affine coordinates $(x_1, x_2)$, is then an affine cone over $\IP^1$.
Thus flat space is an affine cone over complex projective space.

\subsection{A Brief Glossary}
Now that we are working within the realm of projective algebraic geometry, a few concepts liberally used in our analysis should be explained here.
The reader who requires more detail should be able to acquire this from an introductory text such as~\cite{cox,gh,Harris}, or a more advanced one like~\cite{hart}.

\paragraph{Degree:}
We have mentioned the term degree many times.
When the ideal is described by a single polynomial, the degree of the variety is simply the degree of this polynomial.
In the case of multiple polynomials, the degree is the generalization of this, {\em i.e.}\ it is the number of points at which a generic line intersects the variety.

\paragraph{Gr\"obner Basis:}
Any computer package, given an ideal of a set of multivariate polynomials, first places the set in a standard basis.
This Gr\"obner basis is a generalization of Gaussian elimination for a multivariate linear system to general polynomials.
A Gr\"obner basis is defined with respect to some {\bf monomial ordering}, which lets us unambiguously compare two monomials $u = x_1^{\ell_1} \cdots x_n^{\ell_n}$ and $v = x_1^{m_1} \cdots x_n^{m_n}$ and determine when $u < v$.
Given this ordering, $\Gamma$ is a Gr\"obner basis for the ideal $I$ when the ideal given by the leading terms of polynomials in $I$ is itself generated by the leading terms of the basis $\Gamma$.
In computational algebraic geometry, the Gr\"obner basis is determined by (modifications of) Buchberger's algorithm.
Most implementations of the Gr\"obner basis calculation are generically exponential in running time.
This is the main hurdle in performing our computations.
However, for certain problems, lower bounds on the algorithm can be shown to be polynomial~\cite{sgall}.
We hope that the specific nature of our problem may induce a non-prohibitive Gr\"obner reduction.

\paragraph{Hilbert Polynomial:}
This is an extremely important characteristic of a projective variety.
It is {\em not} a topological invariant like the Hodge diamond and depends on the specific embedding.
However, giving an embedding projective space, two varieties with different Hilbert polynomials are clearly distinct.
The polynomial is a generating function of the degree $n$ pieces of the variety.
That is, its $k$-th coefficient is the dimension of the space of degree $k$ homogeneous monomials on the variety.
In particular, the constant is simply the dimension of the projective variety.

\paragraph{Line Bundles:}
We have also used the notation $\CO_{\IP^n}(-k)$ throughout.
This is the line bundle of degree $k$ over $\IP^n$.
In other words, it is a bundle of rank $1$ ({\em i.e.}\ lines fibered over $\IP^n$) whose transition functions are degree $k$ homogeneous polynomials over $\IP^n$.

\newpage
\bibliographystyle{JHEP}

\end{document}